\documentclass[preprint]{aastex63}
\usepackage{multirow}
\usepackage{tablefootnote}
\usepackage{lineno}
\usepackage{soul}
\usepackage{graphicx}
\usepackage{mathtools}
\usepackage{float}
\newtagform{nonums}[\phantom]{}{} 
\usepackage{multibib}
\newcites{meth}{Method References}

\newcommand\Lucy{{\sl Lucy}}

\newcommand\dg{$^{\circ}$}

\newcommand\x{$\times$}

\newcommand\tca{$t_{\rm CA}$}

\newcommand*{\img}[1]{%
    \raisebox{-.3\baselineskip}{%
        \includegraphics[
        height=\baselineskip,
        width=2\baselineskip,
        keepaspectratio,
        ]{#1}%
    }%
}

\published{in {\it Nature}, 29 May 2024}
\graphicspath{{./}{figures/}}

\begin{document}

\setcitestyle{square}

\title{A Contact Binary Satellite of the Asteroid (152830) Dinkinesh}

\author[0000-0001-5847-8099]{Harold F$.$ Levison}
\affiliation{Southwest Research Institute, Boulder, CO, USA}

\author[0000-0003-2548-3291]{Simone Marchi}
\affiliation{Southwest Research Institute, Boulder, CO, USA}

\author[0000-0002-6013-9384]{Keith S.~Noll}
\affiliation{NASA Goddard Spaceflight Center, Greenbelt, MD, USA}

\author[0000-0003-4452-8109]{John R.~Spencer}
\affiliation{Southwest Research Institute, Boulder, CO, USA}

\author[0000-0003-4909-9542]{Thomas S.~Statler}
\affiliation{NASA Headquarters, Washington, DC, USA}

\author[0000-0002-2006-4074]{James F.~Bell III}
\affiliation{Arizona State University, Tempe, AZ, USA}

\author[0000-0001-5890-9821]{Edward B.~Bierhaus}
\affiliation{Lockheed Martin Space, Littleton, CO, USA}

\author[0000-0002-9995-7341]{Richard Binzel}
\affiliation{Massachusetts Institute of Technology, Cambridge, MA, USA}

\author[0000-0002-1804-7814]{William F. Bottke}
\affiliation{Southwest Research Institute, Boulder, CO, USA}

\author[0000-0002-6968-6448]{Daniel Britt}
\affiliation{University of Central Florida, Orlando, FL, USA}

% ...


\vfill
\doublespace
\large
\newpage
\section*{} \label{sec:intro}

%

\newpage
\textbf{Asteroids with diameters less than about 5 km have complex histories because they are small enough for radiative torques (that is, YORP, short the Yarkovsky–O’Keefe–Radzievskii–Paddack effect) \citep{Bottke+2006} to be a notable factor in their evolution \citep{Margot+2015}. (152830) Dinkinesh is a small asteroid orbiting the Sun near the inner edge of the Main Asteroid Belt with a heliocentric semi-major axis of 2.19 AU; its S-type spectrum \citep{Bolin+2023, deLeon+2023} is typical of bodies in this part of the Main Belt \citep{DeMeo&Carry2013}. Here we report observations by the Lucy spacecraft \citep{Levison+2021, Olkin+2021} as it passed within 431 km of Dinkinesh.  Lucy revealed Dinkinesh, which has an effective diameter of only $\sim$720 m, to be unexpectedly complex. Of particular note is the presence of a prominent longitudinal trough overlain by a substantial equatorial ridge, and the discovery of the first confirmed contact binary satellite, now named (152830) Dinkinesh I Selam. Selam consists of two near-equal sized lobes with diameters of $\sim210$ m and $\sim230$ m. It orbits Dinkinesh at a distance of 3.1 km with an orbital period of about 52.7 hr, and is tidally locked.  The dynamical state, angular momentum, and geomorphologic observations of the system lead us to infer that the ridge and trough of Dinkinesh are probably the result of mass failure resulting from spin-up by YORP followed by the partial reaccretion of the shed material. Selam probably accreted from material shed by this event.}

\medskip

Dinkinesh was a late addition to the Lucy mission and was intended primarily as an in-flight test of an autonomous range-finding and tracking system that is a critical component of \Lucy ’s operations \citep{Olkin+2021}.  It was an appealing target because the flyby geometry closely mimicked that of the Trojan targets to be encountered later in the mission \citep{Levison+2021}. \Lucy\ approached Dinkinesh at a solar phase angle of 120°; at close approach the phase dropped rapidly, going through near-zero and then increased to an outbound phase of 60\dg . The relative velocity of \Lucy\ and Dinkinesh was 4.5 km/s. At closest approach, \Lucy\ was $430.629 \pm 0.045$ km from Dinkinesh and had a \Lucy-Dinkinesh-Sun angle of 30$^\circ$. 

A sample of the high-resolution images is shown in Fig.\ref{fig:CAimages}.  The basic shape of Dinkinesh is reminiscent of the `top' shapes seen in the near-earth asteroid (NEA) population (for example, Moshup \citep{Ostro+2006}, Bennu \citep{Barnouin+2019} Ryugu \citep{Watanabe+2019}, and, to a lesser degree, Didymos \citep{Palmer+2023,CampoBagatin+2023}).   Dinkinesh is similarly sized as well.  As described in more detail below, Dinkinesh has an effective diameter of 719 m, while Bennu, Ryugu, and Didymos have effective diameters of between $\sim$560 m and $\sim$900 m.  Like these objects, Dinkinesh is dominated by a prominent equatorial ridge.  Dinkinesh also has a large trough running nearly perpendicular to the ridge.  While both Ryugu and Didymos have similar features \citep{Sugita+2019, Barnouin+2023}, the trough on Dinkinesh appears to be more substantial.  The ridge overlays the trough, implying that it is the younger of the two structures.  However, there is no information on their absolute ages, and thus they could potentially have formed in the same event. 

High-resolution images obtained throughout the encounter (see the Methods/Observations section) make it possible to reconstruct shape models for each of the components. Due to the small size of Dinkinesh and Selam, usefully-resolved imaging was possible for only several minutes before and after close encounter. Dinkinesh's rotation was observed, but the amount of additional terrain revealed by the rotation was small ($\sim\!10\%$) compared to the unilluminated portion of the body. No rotational or orbital motion of Selam was seen. Illumination of Dinkinesh's anti-solar hemisphere from Selam was too faint to be observed. Thus, only one hemisphere of each body is visible in imaging. However, constraints on the unobserved hemispheres can be provided by photometry from both the ground \citep{Mottola+2023} and \Lucy\ when it was too far away to resolve the targets.  We therefore turn our attention to the analysis of this photometry before we further discuss the shapes and structure of the system.

\begin{figure}[H]
    \centering
     \includegraphics[width=6.5in]{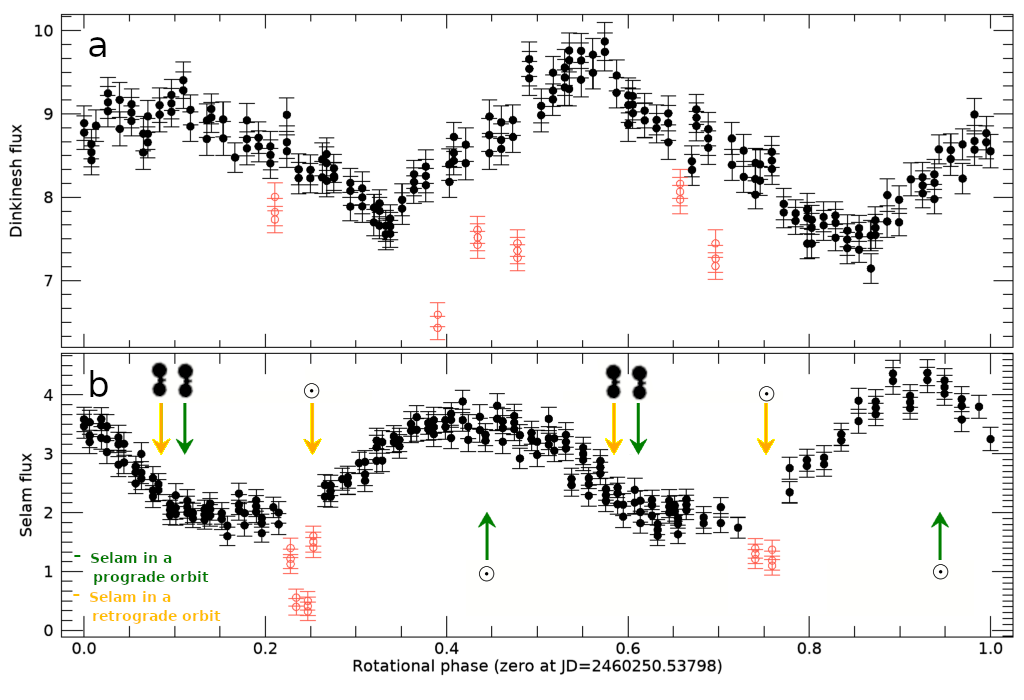}\\
    \caption{{ {\bf Phased lightcurves for Dinkinesh and Selam.} {a) Phased lightcurve for Dinkinesh folded using a period of 3.7387 hr. b) Phased lightcurve for Selam folded using a period of 52.67hr. These periods} were determined from outbound photometry}, as developed in Methods/Lightcurve Analysis section.  The raw photometry is shown in {Extended Data Fig.~\ref{fig:rawlc}}.  The solid black points were used to derive the periods. Dinkinesh's light curve is more complicated than Selam's.  Indeed, Selam's light curve is reminiscent of what is expected for a contact binary consisting of two rotating spheroids seen edge-on and at {this phase angle (60\dg)} \citep{Lacerda2008}. The hollow red points were excluded and correspond to mutual events.  The arrows indicate when different types of events would be predicted.  Events marked with the \Lucy\ spacecraft symbol, \img{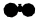}, show occultations (when one object passes in front of the other from the spacecrafts point-of-view) if they occur.  Events marked with the sun symbol, $\odot$, indicate the potential times of eclipses (where the shadow of one object falls on the other). The observed mutual events are associated with eclipses. Occultations are not seen by \Lucy\ during departure, which is consistent with the fact that its trajectory is slightly inclined with respect to Dinkinesh's orbital plane.   {Green} arrows show events that occur if Selam were in a prograde orbit about Dinkinesh, while {orange} arrows occur for a retrograde orbit.  From this we can conclude that Selam's orbit is retrograde.}
    \label{fig:DSlc}
\end{figure}

The unresolved data from the post-encounter light curve photometry campaign (see the Methods/Observations section) is described in Fig.~\ref{fig:DSlc}.  From these data we determine that the contribution of Selam's rotation to the lightcurve has periodicty with T = 52.44 $\pm$ 0.14 hrs, comparable to the 52.67 $\pm$ 0.04 hrs period found from ground-based observations \citep{Mottola+2023}. We adopt the ground-based period of 52.67 hrs because it is more precise due to its longer sampling baseline. The post-encounter light curve also shows dips inferred to be due to mutual eclipses of Dinkinesh and Selam with the same 52 hour periodicity (Fig.~\ref{fig:DSlc}, and the Methods/Observations section), demonstrating that Selam's orbital period is very similar to its rotational period.  We interpret this to mean that the system is tidally locked.  By using the formalism in \citep{Jacobson&Scheeres2011a}, we estimate that the timescale for tidal effects to align the long axis of Selam radially relative to Dinkinesh to be short, of order $10^{5}$ yr at the current separation, although their formalism might not be accurate because{ some important radiation effects \citep{Bottke+2006}} were not considered \citep{Pou+Nimmo2024}.  We also find that the centers of Dinkinesh and the two lobes of Selam {appear to} lie along a single line ({Fig.\ref{fig:CAimages}m}) --- consistent with a tidally locked system. Thus, we conclude that Selam is in synchronous rotation and thereby orbits Dinkinesh with a period of 52.67 hrs.  The timing of the mutual events in the post-encounter lightcurve (Fig.~\ref{fig:DSlc}), relative to Selam's orbital position during the flyby, shows that Selam's orbit must be retrograde with respect to Dinkinesh's heliocentric orbit.

The primary, Dinkinesh, rotates more rapidly, with the best-fit to the lightcurve giving a spin period of P = 3.7387 $\pm$ 0.0013 hr. Feature tracking during the flyby shows that the rotation is retrograde with respect to ecliptic North, i.e., in the same sense as Selam's orbit. The overall spin state (a synchronous secondary and a rapidly spinning primary) makes Dinkinesh similar to the majority of small near-earth and Main Belt asteroids with close satellites \citep{Pravec+2016}.

We now return to the topic of the {shapes of Dinkinesh and Selam. A model of Dinkinesh} produced by the process described in the Methods/Shape section and based on a preliminary reconstruction of \Lucy's trajectory is illustrated in Fig.~\ref{fig:ShapeModel}. We find a volume-equivalent spherical diameter of 719 $\pm$ 24 m for Dinkinesh based on this shape model{.  Selam appears to consist of two distinct lobes.  However, the contact point was in shadow during the encounter and so the exact nature of the neck is uncertain.  Images taken during approach where the outer lobe was father way from the spacecraft than the inner one (see {Fig.~\ref{fig:CAimages}h} for example) show that the neck is less than $\sim\!67\%$ of the inner lobe's diameter.   We find} equivalent spherical diameters of 212 $\pm$ 21 m and 234 $\pm$ 23 m for the inner and outer lobes of Selam based on fitting ellipses to visual limb profiles.  {If the lobes were in orbit about one another, their period would be $\sim\!4$ hr, which is inconsistent with the lightcurve observations described above.  Additionally, we would have detected motion if the period were that short.  Thus, the lobes must be resting on one another and Selam is likely a contact binary.}

{Outbound images clearly show both lobes of Selam ({Fig.\ref{fig:CAimages}m}) from a direction almost perpendicular to the vector between them, as determined by triangulation.  From these images, we derive a preliminary estimate of the center-of-figure separation between Dinkinesh and Selam to be $3.11 \pm0.05$ km at the time of the flyby.  We argue in the Methods/Mass\&Density section that Selam is on a circular orbit. If so, this separation represents the semi-major axis of the mutual orbit.}

The orientation in space of both Dinkinesh and Selam can be estimated with current data.  In particular for Dinkinesh, the small amount of rotation observed during the encounter and the direction of its shape model's short axis suggests that its obliquity {$\sim\!178.7 \pm 0.5$\dg\ (i.e. its rotational axis is $\sim\!1$\dg\ from} being perpendicular to its orbital plane). For the satellite, the mutual eclipses observed during the post-encounter Lucy observations, and mutual events inferred from the 2022--2023 ground-based lightcurve [6], suggest that its orbit plane is close to Dinkinesh's heliocentric orbital plane.   It is therefore likely that all three, Dinkinesh's heliocentric orbit, Selam's orbit, and Dinkinesh's equatorial plane are close to one another. This configuration is nearly ubiquitous among small binary asteroids {\citep{Margot+2002}} as a result of spin-pole reorientation by {the asymmetric thermal radiation forces caused by the YORP effect \citep{Bottke+2006}}.  The YORP timescale is less than $\sim 10^7$ yr for Dinkinesh's spin-pole to approach either zero or 180\dg\  \citep{Pravec+2012, Statler2015}.

The inner lobe of Selam also has a prominent ridge-like structure ({Fig.~\ref{fig:CAimages}h-k}). Both lobes of Selam have flat facets and a blocky, angular overall shape, and the apparent ridge may be the boundary of two such facets. If, however, the structure formed from the accretion of material from a Dinkinesh-centered disk, as one might expect, it would have originally been aligned with both the orbit plane and Dinkinesh's ridge. {In that case, it} is likely that Selam's ridge then became misaligned during the formation of the contact binary{, but this implies that either 1) Selam is currently rotating or librating about its long axis, or 2) its ridge formed before contact.}{ The observed structure of Selam implies that it is a rubble pile, at least partially.  However, the} angular, binary, shape of Selam implies significant internal strength, and is dramatically different from the oblate spheroid shape of Dimorphos, the moon of Didymos \citep{Daly+2023} the only other satellite of a sub-km asteroid (also an S-type) for which we have detailed images.

The mineralogy and bulk density of Dinkinesh provide constraints on its structure. Dinkinesh's bulk density is {$2400 \pm 350$ kg/m$^3$ (Methods/Mass\&Density section), which is} in the range of expected values for objects with ordinary chondrite mineralogies. Bulk densities of L-chondrite meteorites, which {are a good analog for the range of ordinary chondrites \citep{Scheeres+2015} and have the expected mineralogy for S-type asteroids,} average $3360 \pm 160$ kg/m$^3$ with 7.5\% microporosity \citep{Macke2010}. {Given the uncertainties in Dinkinesh bulk density discussed in the Methods/Mass\&Density section this suggests a macroporsity of $25\!\pm\!10\%$.} Its bulk density is in family with the S-type NEAs of this mineralogy and in this size range. For example, while Didymos has a similar density of $2800 \pm 280$ kg/m$^3$ \citep{Barnouin+2023}, Itokawa’s bulk density is $1900 \pm 130$ kg/m$^3$ \citep{Fujiwara+2006} and the radar-observed binary Moshup is $1970 \pm 240$ kg/m$^3$ \citep{Ostro+2006}. The low-density objects are likely much more porous and have a more pronounced rubble-pile structure than Didymos, with Dinkinesh some place in between. Dinkinesh and Didymos are probably on part of a continuum of where significant portions of the object are relatively coherent.

Dinkinesh accounts for 94\% of the volume of the system with Selam accounting for 6\%. If we assume that all of the components have an equal density, the component masses of Dinkinesh and Selam are M$_D = 4.67 \times 10^{11}$ and M$_S = 0.28 \times 10^{11}$ kg, respectively. Using these component masses, it is possible to calculate that the barycenter is offset from the center of mass of Dinkinesh by a distance $s_{bary} = 176$ m in the direction of Selam, well interior to the body of the primary.

Fig.\ref{fig:CAimages} strongly suggests that Dinkinesh suffered a global structural failure in its past.  Given its small size, this event is likely the result of spin-up by the YORP effect \citep{Hirabayashi+2015,Bottke+2006}{, see discussion in Fig.~\ref{fig:Cartoon} caption}.  If true, then the Dinkinesh system's angular momentum should be comparable to the total angular momentum of a parent body spinning near the spin-barrier limit \citep{Pravec+Harris2007}.  Indeed, we find that the Dinkinesh system contains 88\% of the angular momentum required for rotational breakup (cf$.$ Methods/Angular$\_$Momentum section), which is consistent with the idea that Dinkinesh's structure failed due to its large angular momentum.

Dinkinesh shares many characteristics with other similar-sized asteroids, both Near Earth and Main Belt and is the only sub-km size Main Belt object ever studied at close range. Approximately 15\% of small asteroids are observed to be binaries \citep{Margot+2002, Pravec+2006}. For the subset of these systems that are well characterized, the dominant pattern is a system with a synchronous secondary in a near-circular orbit with a semi-major axis, $a$, of $\approx 3$ or more primary radii, $r_{prim}$ \citep{Pravec+2006}. Selam's semi-major axis, at $a/r_{prim} \approx 9$, is wide compared to the majority of other well-characterized systems of similar size that cluster closer to $a/r_{prim} \approx 3$ \citep{Monteiro+2023}. Dinkinesh's spin period is also longer than the $\approx 2.5$ hr period typically observed in the NEA binary population \citep{Pravec+2006}. One possible scenario is that Selam originally formed nearer to Dinkinesh and then evolved to a larger semimajor axis through tidal interaction and/or binary YORP that also slowed down Dinkinesh's rotation \citep{Jacobson&Scheeres2011b}.  

The most distinctive characteristic of the Dinkinesh - Selam system is the contact binary structure of Selam. Fig.\ref{fig:Cartoon} illustrates three possible scenarios for its formation.  The binary nature of Selam places important constraints on the formation of these satellite systems no matter how it formed.  First, the fact that the two lobes are nearly the same diameter {argues} that the satellite formation process responsible for Selam {favors} building objects of a particular size. As far as we are aware, none of the formation models in the literature has been shown to meet this requirement.  Second, {as we describe above}, the two lobes are distinct bodies.  So, the process that brought the two lobes together must have done so with a small enough velocity for the lobes to have survived.  

The unexpected complexity of the Dinkinesh system strongly suggests that small asteroids in the Main Belt are more complex than previously thought.  The fact that a contact binary can form in orbit about a larger object suggests a new mode for the formation of small bilobed bodies such as Itokawa{\citep{Fujiwara+2006}}, for which they {may once have been} components of a system such as Dinkinesh that subsequently became unbound.

\bibliography{DinkySelam}

\newpage
\section*{\large\bf Figure Captions}

\includegraphics[width=7.0in]{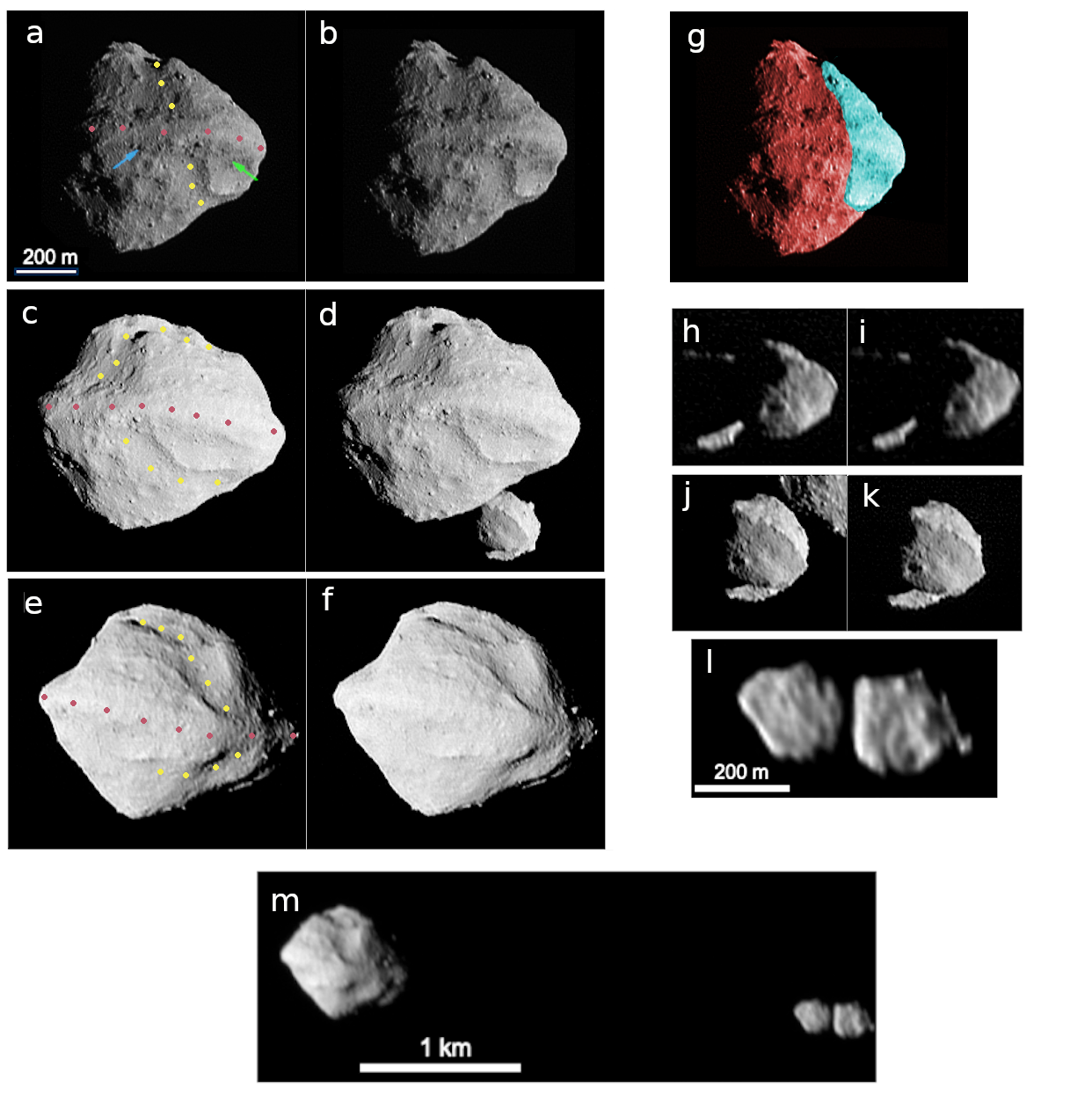}
\begin{figure}[H]
    \centering
\caption{{ {\bf Images of Dinkinesh and Selam obtained by \Lucy's close approach imaging campaign.}  {a--f})  Cross-eyed stereo versions of the images taken on approach, near close approach, and on departure, respectively (see the Methods/Observations section for description of imaging campaign).} Dinkinesh has two major geological features: a longitudinal trough and an equatorial ridge (the yellow and rose colored dots, respectively). The two colored arrows (green and blue) in {Fig.~\ref{fig:CAimages}a} point to the northern boundary of the ridge, as determined by visual inspection, at the two fiducial longitudes discussed in Fig.~\ref{fig:ShapeModel}.  {g}) A simulated image of Dinkinesh with the trough removed.  This is a modified version of {Fig.~\ref{fig:CAimages}a} where the cyan region was moved 79 m to the lower left in the image (26$^\circ$ from horizontal) and rotated 7$^\circ$ clockwise.  We take the fact that the limb profile of Dinkinesh is smooth near the color transitions of this reconstruction to suggest that the trough is a result of a structural failure that moved the cyan region away from the remainder of the body.  {h--k}) Stereo pairs {of images} of Selam taken on approach and near close approach, respectively. {l}) A single image of Selam taken on departure.  Selam was outside of the L'LORRI field-of-view from 10 sec to 5.5 min after close approach and so stereo imaging is not possible. The images of Selam allow us to visually estimate the dimensions of its lobes by crudely approximating their complex shapes as triaxial ellipsoids.  We find that the inner and outer lobe major axes lengths in the directions parallel to the Dinkinesh vector, the orbital direction, and the spin pole are roughly $240 \times 200 \times 200$m and $280 \times 220 \times 210$m, respectively.  {m}) A departure image of the entire system. Also, all images are deconvolved except {Fig.~\ref{fig:CAimages}m. {Fig.~\ref{fig:CAimages}l} and m} are the same image, so comparison illustrates the effects of deconvolution. Ecliptic north is approximately up in all frames{, while  Dinkinesh's body north is down because it is a retrograde rotator.} Image details are as follows.  Times relative to close approach in minutes: {a: -1.04, b: -1.29; c: +0.21, b: -0.04; e: +2.21, f: +1.71; h: -2.29, i: -3.29; j: -0.29, k: -0.54; l \& m: +5.46.   Original pixel scale, m/pixel: a: 2.53, b: 2.72; c: 2.14, d: 2.12; e: 3.63, f: 3.11; h: 3.70, i: 4.85; j: 2.16, k: 2.24; l \& m: 7.56.  Solar phase angle, degrees: a: 62.1, b: 68.0; c: 21.5, d: 30.5; e: 25.0, f: 17.8; h: 84.2, i: 93.3; j: 39.3, k: 47.7; l \& m:} 44.5.}
\label{fig:CAimages}
\end{figure}

\begin{figure}[H]
    \centering
     \includegraphics[width=6in]{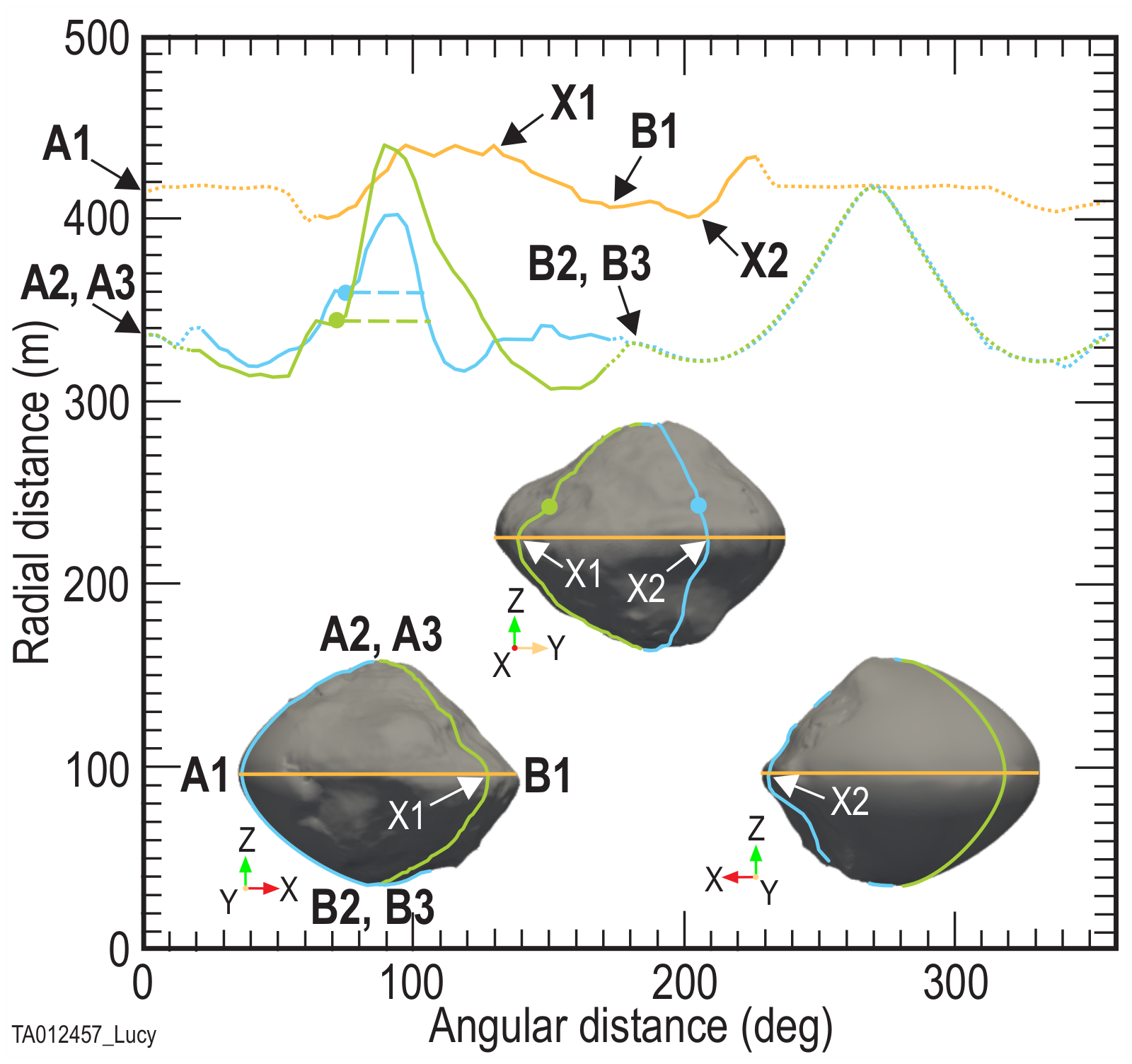}\\
\caption{{ {\bf Dinkinesh’s shape model.} Three orientations (insets) and} topographic cuts designed to emphasize the
structure of the equatorial ridge. The model, which is described in the Methods/Shape section, consists of two regions.  The side of the model that was facing \Lucy\ during the encounter was based on the images taken during close approach imaging campaign (see the Methods/Observations section). The unilluminated portion of Dinkinesh is estimated with a super-ellipsoid. {The rotational ($z$) axis points up in these figures.} The orange curve shows a latitudinal cut that lies along the ridge. The blue and green curves are longitudinal cuts corresponding to minimum and maximum elevation of the equatorial ridge, respectively.  The points labeled X1 and X2 indicate where the green and blue curves cross the orange curve, respectively.  The large dots show the location of the ridge's northern boundary, as determine visually in {Fig.\ref{fig:CAimages}a} (the corresponding blue and green arrows), and the horizontal dashed lines show the extent of the ridge. The ridge at its lowest point measures 150 m wide and 40 m high, while it is 230 m wide and 100 m high at its highest point.  The curves are solid in the locations where the shape model is reliable, while they are dotted elsewhere.  The reference locations labeled A1, A2, A3, B1, B2, B3, X1, and X2 are included to allow the reader to associate the shape model to the curves.}
\label{fig:ShapeModel}
\end{figure}

\begin{figure}[H]
    \centering
    \includegraphics[width=7in]{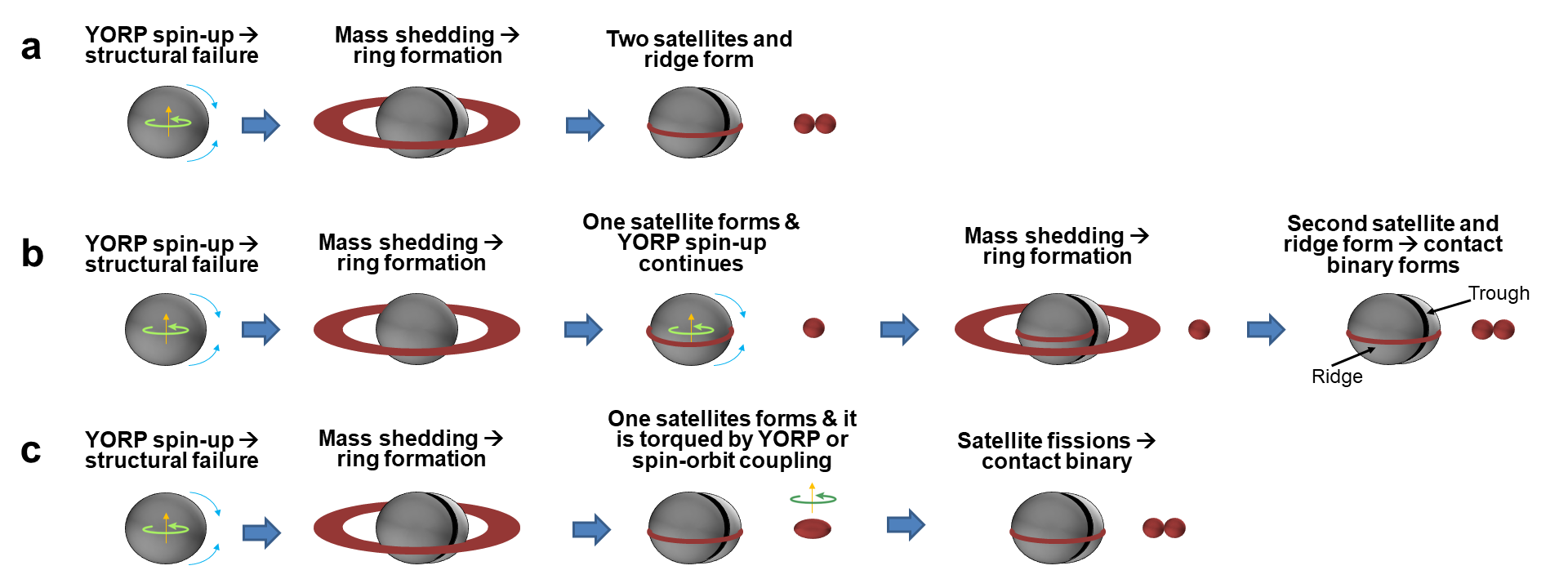}\\
\caption{{ {\bf A simplified graphical depiction of a plausible sequence of events leading to the current configuration of the Dinkinesh-Selam system.}} Asteroids with diameters less than $\approx$ 10 km are subject to spin-up by the YORP effect \citep{Bottke+2006}. Rapid spin of the primary and the associated centrifugal force eventually trigger a structural failure that leads to sudden mass shedding \citep{Hirabayashi+2015}. This event might also have created the trough seen on Dinkinesh ({Fig.\ref{fig:CAimages}a--f}) through the mass movement of a portion of the body ({Fig.\ref{fig:CAimages}g}). The shed material forms a ring with some material coalescing into a satellite(s) and closer material eventually falling back to the surface at the equator to form the ridge \citep{Hyodo+2022}. The formation of the contact binary may be the result of a merger of two satellites formed either {a)} in a single mass-shedding event, or {b)}in two separate events \citep{Madeira+2024}. {c)} An alternative scenario is that Selam formed as a single object that subsequently underwent fission due to spin-orbit coupling \citep{Jacobson&Scheeres2011a, Jacobson+2016} . {It is also possible that some or all of Selam formed from a collision on the primary \citep{Michel+2020}, but the trough and ridge would not have survived such an event.  Thus, this mechanism would still require later Dinkinesh spin up by YORP and mass shedding to form the trough and superposed equatorial ridge.}}
\label{fig:Cartoon}
\end{figure}

\newpage
\section*{\large\bf METHODS} \label{sec:Methods}

\noindent{\bf Observations:} 

The analysis presented here is based on panchromatic (350 --- 850 nm) images taken with \Lucy's LOng Range Reconnaissance Imager, hereafter L'LORRI, which is a 20.8 cm, f/13 telescope feeding a 1024 $\times$ 1024 pixel CCD focal plane \citep{Weaver+2023}. L'LORRI has a field of view of 0.29$^\circ$ and a pixel size of 5 $\mu$rad.  L'LORRI was primarily used in three distinct observation campaigns during the encounter:  1) Optical navigation reconstruction images were designed to {precisely determine} \Lucy's trajectory.  They were taken daily during the period of $\pm 4$ days of encounter (\tca = --4 to +4 days) and every 15 minutes from \tca = --2 hr to +2 hr. 2) High-resolution close approach images, which were taken every 15 seconds from \tca = --10 min to +9 minutes{, then with 1 minute cadence until +55 minutes}.  3) Post-encounter light curve photometry was acquired from \tca = +4 hr to +95 hr. {Three exposures} were taken at a cadence of one hour.  At this time the {Dinkinesh-Selam} system was unresolved.  In order to minimize data volume, these data were taken in L'LORRI's so-called 4$\times$4 mode, which bins the data by 4$\times$4 pixels during the CCD read out.

\medskip
\noindent {\bf Lightcurve\_Analysis:} 

The orbital period of Selam and the rotational period of Dinkinesh can be determined using the post-encounter light curve photometry described above in the Methods/Observations section. Instrumental magnitudes of the system were extracted from the images using a 1.5-pixel radius aperture.  The small aperture served to exclude contamination from nearby stars.  The formal errors from the extraction were scaled upward by a factor of 1.545 to adjust the reduced $\chi^2$ to be 1 prior to determining the final uncertainties on the fitted results.  There were 267 images analysed.

The data were compensated for the changing distance as well as correcting to a constant solar phase angle using a phase coefficient of 0.06 mag/deg.  The phase angle varied from 60.52\dg\ at the start to 59.67\dg\ at the end.  The observing direction changed little over the 3.5 days and these corrections remove these slight changes leaving only a record of the global photometric properties of the system. The resulting lightcurve is shown in { Extended Data Fig.~\ref{fig:rawlc}} in units of relative flux.

We analysed the lightcurve with an iterative process designed to separate the contributions to the total flux from Dinkinesh and Selam. As the first step a model was constructed that consisted of a Fourier series expansion of the lightcurve combined with a period for each object.  The reference time for the rotational phase was arbitrarily set to the time of the first data point for both objects.  The mean flux of Dinkenesh was a free parameter in the model.  {In addition, we iteratively varied the Selam/Dinkinesh mean flux ratio.  This ratio is constrained by the close-approach resolved images {(Fig.~\ref{fig:CAimages}d}, for example) which show that the ratio of the visible areas of the two objects is 0.25.  The two objects are also seen to have similar surface brightness, and so the unresolved flux ratio is also 0.25.  This ratio was assumed to be at minimum light for both objects because Selam is viewed edge-on.}  An iterative correction was applied after separating the lightcurves to correct from the minimum to the mean flux and the final mean flux ratio was set at 0.33 (corresponding to a magnitude difference of 1.3).

The model parameters were determined in a series of iterative steps.  The first pass fit set a reasonable mean flux for Dinkinesh and the Fourier terms were disabled.  At this point, only Selam was free to be adjusted to fit the data.  The data were scanned in period.  At each step, a best fit Fourier series was computed and the $\chi^2$ was recorded. The lowest $\chi^2$ period gave a preliminary value of 51.76 hours for Selam.  This model was subtracted from the lightcurve data and a similar scan was performed on the Dinkinesh-only data. The Dinkinesh scan returned two interesting minima in $\chi^2$ at periods of $\sim$3.7 and $\sim$4.3 hours. Note that all periods assume that the lightcurve is double-peaked.

Given the two preliminary periods, the data were then fitted with the full model from the two objects and all free parameters were optimized simultaneously with an amoeba $\chi^2$ minimization \citep[][Chapter 10.4]{Press+1992}.  Using the amoeba fit as the starting point with the {\it a posteriori} correction to the uncertainties, a second Markov-Chain Monte-Carlo fit \citep[cf.][]{Foreman-Mackey+2013} was run for the model.  There were 18 data points that were excluded due to unreasonably large residuals (see the discussion below). The final fitted lightcurves revealed amplitudes of 0.82 mag for Selam and 0.25 mag for Dinkinesh.

The Selam rotation period was determined to be 52.44 $\pm$ 0.14 hours from this fit, but it is also attributed to its orbital period about Dinkinesh because it is likely tidally locked, as shown by the presence of mutual events. The resulting phased lightcurves are shown in Fig.~\ref{fig:DSlc}.

The variation in flux for the two objects coincidentally are about the same. Dinkinesh is much larger, which implies that it has a smaller relative variation in its flux.  The lightcurve of Selam is well fit by two Fourier terms that capture the slightly asymmetric maximum and slightly broadened minima.  The lightcurve of Dinkinesh is considerably more complicated, both the minima and maxima are asymmetric, but there are also clearly higher order variations seen.  In this case a 4-term Fourier fit was required and even this does not fully capture all of the detail in the curve.  For instance, one of the minima is sharper than can be followed with a 4-term fit. {Dinkinesh's rotation period was determined to be 3.7387 $\pm$ 0.0013 hr (the 4.3 hour period discussed above was determined to be an alias).}

The outliers that were flagged during the lightcurve fitting, which are shown in red in the figures,  are also of interest because they occur at a coherent rotation phase following a similar time after the two lightcurve minima for Selam.   A reasonable explanation for these low points is a mutual event between the two bodies. These could, in general, be from the bodies occulting each other from the spacecraft's perspective, or from casting shadows on one another.   {Fortunately, the timing of these minima allows us to determine which.}

Looking at the photometry as a function of time, the low points appear at a regular interval at half the rotation period of Selam.  Geometric constraints from the absolute timing indicate that the events are shadow transits of each other and not physical obscuraton along the line of sight (occultations).  Furthermore, the timing clearly indicates that the orbital motion of Selam is retrograde, as is true for the rotation of Dinkinesh as well.  The first and third dips seen in time are inferior shadowing events while the middle dip is a superior event.  In the phased plot, the two inferior events overlay each other and trace out a more complete lightcurve of an event.  The superior event has fewer measurements and shows an incomplete profile of the dip that misses the maximum eclipse point that must be in the middle between the two sets of points. 

\medskip
\noindent {\bf Shape:} 

The digital shape model used for this study (see Fig.\ref{fig:ShapeModel}) was generated by applying classical stereo-photogrammetry techniques \citep[][and references therein]{Preusker+2017} to L’LORRI imagery. A total of 48 images with a best ground sampling distance ranging from $\sim$10 m/pix to 2.2 m/pix were chosen from the high-resolution close approach images described in the Methods/Observations section.  These were employed to establish a network of ~3000 control points, which served as an input for the bundle adjustment process. {Further, thanks to the very good noise and sensitivity
performance of the L'LORRI imager, and to its comparatively large field of view, we could identify about 20 catalog field stars in the Dinkinesh fields throughout the encounter. These star positions were used in the determination of the stereo-photogrammetric adjustment, and contributed considerably to stabilize the solution.}

As a result, the camera extrinsic matrices were determined, which describe the transformation between the camera’s and the body-fixed reference system. These transformation matrices were then used to triangulate surface points from homologous image points, which were derived by means of dense stereo-matching \citep{Wewel1996}. The resulting dense point cloud ($\sim 5 \times 10^6$ 3D points) was then connected into a regular triangular mesh. The shape model derived from stereo reconstruction {has an estimated scale error of about 1.4\%, and covers} about 45\% of the body’s surface. In order to produce a closed shape, and allow an estimation of the body volume, the unseen hemisphere has been approximated with an analytical solid figure. For this purpose, we chose a generalized super-ellipsoid \citep{Ni+2016}, whose implicit representation is given by the function
\usetagform{nonums}\begin{equation} 
1 = \left| \frac{x}{a} \right|^k + \left| \frac{y}{b} \right|^m + \left| \frac{z}{c} \right|^n,
\end{equation}
where $x$, $y$, and $z$ are the standard Cartesian coordinates.  A fit to the reconstructed hemisphere leads to $a = 0.40$, $b = 0.40$, $c = 0.35$ km, $k = m = 2$, and $n = 1.35$.  The generalized super-ellipsoid provides a better match to the Dinkinesh’s `top' shape than a conventional triaxial ellipsoid.

We estimated the uncertainty in Dinkinesh's volume from the difference between the shape model and the super-ellipsoid convex shell. For the hemisphere covered by imaging, the difference in volume is 4.7\%. {In order to be conservative, we round this and apply an arbitrary} factor of two margin to arrive at the volume uncertainty of $\pm$ 10\%. This uncertainty is propagated to quantities derived from the volume. In particular, we note that the volume equivalent radius of Dinkinesh is calculated as $r_{veq} = (3V/4\pi)^{1/3}$ rather than from direct distance measurements. 

The dimensions of the two lobes of Selam were found by fitting ellipses to orthogonal axes in multiple resolved images of Selam from different viewing angles. Selam's inner lobe is fit with an ellipsoid measuring $240 \times 200 \times 200$ m. The outer lobe is measured at $280 \times 220 \times 210$ m. {Uncertainties were estimated to be 10\% per axis by adjusting the ellipsoidal fits until they were visually too large or too small to match the images.}
 
Combining the above values, we calculate a total system volume of $V_{tot}=2.06 \pm 0.20 \times 10^8$ m$^3$.

\medskip
\noindent {\bf Mass\&Density:} 

System density can be estimated from the orbital period and {relative semi-major axis} of the two bodies. {As we describe in the main text,  the center-of-figure separation between Dinkinesh and Selam was $3.11 \pm$ 0.05 km at the time of the flyby.}  The eccentricity of Selam's orbit is not directly derivable from existing data, although it can be constrained. The regular phasing of the lightcurve minima collected prior to encounter from the ground \citep{Mottola+2023} and from \Lucy\ (Fig.\ref{fig:DSlc}) is consistent with a near-circular orbit, given our inference (Fig.~\ref{fig:DSlc}) that these minima are caused by mutual eclipses. {We would expect Selam’s eccentricity to be near zero given that tidal timescales for orbit circularization} are of order 10$^6$ - 10$^7$ yrs. The ages of asteroid pairs where one of the members of the pair has subsequently undergone a mass-shedding event leading to the formation of a satellite suggest that binary-YORP effects \citep{Cuk+Burns2005} might shorten the circularization timescale to less than $\sim 10^6$ yrs \citep{Pravec+2019,Pou+Nimmo2024}.  Thus, we assume $e = 0$ in the analysis {performed here}. Ground-based lightcurve observations, taken at multiple epochs, can better constrain any orbital eccentricity that might exist.

Assuming that Selam is in a circular orbit about Dinkinesh and has an orbital period of $52.67 \pm 0.04$ hr, we derive a system mass of $4.95 \pm 0.25 \times 10^{11}$ kg ($GM = 33.0 \pm 1.6\ {\rm m}^3/{\rm s}^2$) from Kepler's third law.   {In the Methods/Shape section, we calculate a total system volume} of $V_{tot}=2.06 \pm 0.20 \times 10^8$ m$^3$. Combining the system mass and volume we derive a bulk density of $\rho = 2400 \pm 350$ kg/m$^3$.  We add the caveat that if the assumption of zero eccentricity is incorrect and the separation observed at the time of the flyby differs from the semi-major axis, it would introduce a systematic error into the calculation of density. Conversely, however, the range of likely density for a S-type asteroid, as discussed below, constrains the maximum eccentricity to be of order of 0.1 and the assumption of zero eccentricity is fully consistent with known asteroid properties.

\medskip
\noindent {\bf Angular$\_$Momentum:} 

Knowledge of the component masses and the spin state can be combined to calculate the system's angular momentum. For simplicity we assume that Dinkinesh's moment of inertia can be adequately represented by a sphere of volume-equivalent radius. Assuming that Selam is tidally locked, the contribution to the angular momentum from its spin is small. Likewise, the orbital motion of Dinkinesh around the barycenter is small and we ignore it. The system angular momentum is nearly equally divided between Dinkinesh's spin, $L_{spin} = 11.2 \pm 1.9 \times 10^{12}$ kg m$^2$ s$^{-1}$, and Selam's orbital motion, $L_{orb} = 8.0 \pm 4.0 \times 10^{12}$ kg m$^2$ s$^{-1}$. The total angular momentum of the system is $L_{sys} = 19.3 \pm 4.4 \times 10^{12}$ kg m$^2$ s$^{-1}$.
The normalized angular momentum, $\alpha_L$, is computed from the total system angular momentum divided by the angular momentum of a sphere containing the total mass of the system rotating at the maximum rate for a cohesionless rubble pile \citep{Pravec+Harris2000}. That rate is given by $\omega_{max} = (4 \pi \rho G / 3)^{1/2}$, corresponding to a spin period of $T_{max} = 2.13$ hrs, i.e., the observed Main Belt spin barrier. We find $\alpha_L = 0.88$, consistent with that expected for a binary produced by fission \citep{Pravec+Harris2007}.

\newpage

\medskip\centerline{\bf END NOTES}

\smallskip\noindent{\bf Acknowledgements}\ \  

The \Lucy\ mission is funded through the NASA Discovery program on contract No.~NNM16AA08C.  The authors thank the entire \Lucy\ mission team for their hard work and dedication.

\medskip\noindent{\bf Author Contributions}\ \ 

H.F.L. and K.S.N. jointly led the writing of the text. S.Mo., F.P., and S.Ma. developed the shape model. J.R.S., I.S., J.S., and S.Ma. planned the science encounter sequence. J.R.S. led the production and analysis of L'LORRI images. M.B., J.R.S, and S.Mo. analyzed Lucy lightcurve data to derive Selam's orbital period and mutual event timing. T.R.L contributed to deconvolution of L'LORRI images. B.H.M. led the production of stereo images from deconvolved L'LORRI images. R.M. identified Dinkinesh as a possible target for \Lucy. All of the authors contributed to \Lucy\ science, science planning and the successful operation of \Lucy\ before, during, and after the encounter.

\medskip\noindent{\bf Competing Interest Declaration}

The authors declare no competing interests.

\medskip\noindent{\bf Corresponding Author}

Correspondence and requests for materials should be addressed to Harold~F.~Levison, Southwest Research Institute, Boulder, CO.  hal.levison@swri.org

\medskip\noindent{\bf Data Availability Statement}

All raw and calibrated images, as well as the digital shape model, will be available via the Planetary Data System (PDS) (https://pds-smallbodies.astro.umd.edu/data\_sb/missions/lucy/index.shtml) by 31 August 2024.

\newpage
\section*{\bf\large EXTENDED DATA}

\renewcommand{\figurename}{{Extended Data Figure}}
\setcounter{figure}{0}
\begin{figure}[H]
    \centering
        \includegraphics[width=1.0\textwidth,trim={0 7.5cm 0 7.0cm},clip]{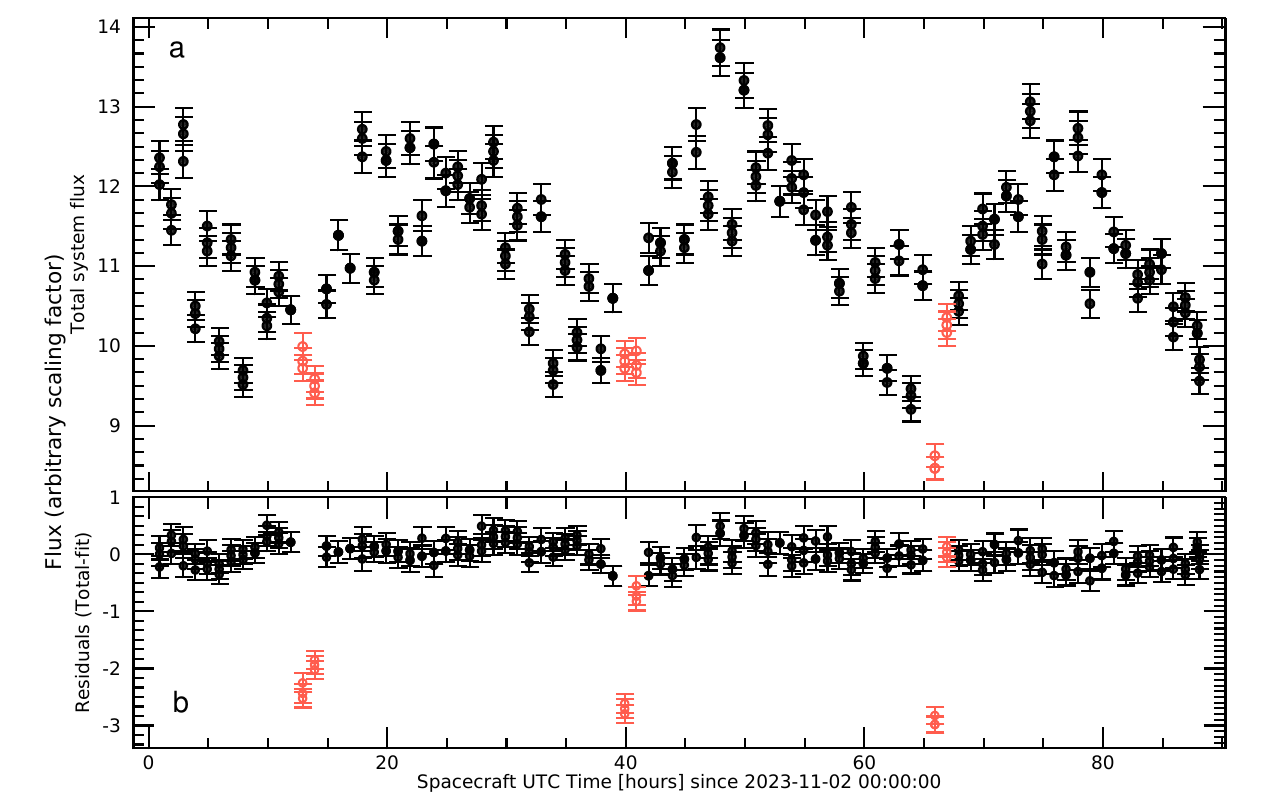}
    \caption{{ \bf Post-encounter photometry of the Dinkenesh/Selam system.}  {a) The observed flux (with arbitrary scale) of the system as a funtion of time.  b) The residuals to the fit described in the Methods/Lightcurve\_Analysis section.}  The solid black points are the data used in the combined lightcurve fit.  The hollow red points are those that were excluded from the fit due to being affected by mutual events between the components.
    }
    \label{fig:rawlc}
\end{figure}

\end{document}